\documentclass[12pt]{article}
\usepackage{amsmath}
  {
	\newtheorem{assumption}{Assumption}
}
\usepackage{amsfonts}
\usepackage{graphicx,psfrag,epsf}
\usepackage{enumerate}
\usepackage[round]{natbib}
\usepackage{url} 
\usepackage[linesnumbered,ruled]{algorithm2e} 
\usepackage{multirow}

\newcommand{\blind}{0}

\begin{document}
	
\newcommand{\Perp}{\mathrel{\text{\scalebox{1.07}{$\perp\mkern-10mu\perp$}}}}

\def\spacingset#1{\renewcommand{\baselinestretch}%
{#1}\small\normalsize} \spacingset{1}


\if0\blind
{
  \title{\bf Tree-Based Bayesian Treatment Effect Analysis}
  \author{Pedro Henrique Filipini dos Santos 
	\hspace{.2cm}\\
    Institute of Mathematics and Statistics, University of S\~ao Paulo\\
    and \\
    Hedibert Freitas Lopes \\
    INSPER Institute of Education and Research}
  \maketitle
} \fi

\if1\blind
{
  \bigskip
  \bigskip
  \bigskip
  \begin{center}
    {\LARGE\bf Tree-Based Bayesian Treatment Effect Analysis}
\end{center}
  \medskip
} \fi

\bigskip
\begin{abstract}
The inclusion of the propensity score as a covariate in Bayesian regression trees for causal inference can reduce the bias in treatment effect estimations, which occurs due to the regularization-induced confounding phenomenon. This study advocates for the use of the propensity score by evaluating it under a full-Bayesian variable selection setting, and the use of Individual Conditional Expectation Plots, which is a graphical tool that can improve treatment effect analysis on tree-based Bayesian models and others ``black box'' models. The first one, even if poorly estimated, can lead to bias reduction on the estimated treatment effects, while the latter can be used to found groups of individuals which have different responses to the applied treatment, and analyze the impact of each variable in the estimated treatment effect. 
\end{abstract}

\noindent%
\vfill

\newpage
\spacingset{1.45} 
\section{Introduction}
\label{sec:intro}

The advent of Bayesian computation on the previous decades allowed the creation of models with a high degree of complexity, while Monte Carlo Markov Chains (MCMC) algorithms allow the estimation of models that were previously considered infeasible. One of these models is the Bayesian Additive Regression Trees (BART) \citep{CGM10}, an extremely flexible nonparametric model. \cite{Hill11} applied the BART models to the causal inference setting, more specifically on the estimations of binary treatment effects for observational studies, and the results were promising. 

But these models can be affected by the regularization-induced confounding \citep{Hahn18}, which states that, in the presence of confounding, the regularization of these models may cause biased estimations of the treatment effects. \cite{Hahn17} argue that, under strong ignorability, this problem can be eased through the use of the propensity score \citep{Rubin83} among the covariates of the model.

This study has two main contributions. The first is the application of a sensitivity analysis in the causal inference setting through the use of Individual Conditional Expectation Plots \citep{Goldstein15}. The second is to corroborate the inclusion of the propensity score on Bayesian regression tree models \citep{Hahn17} by using simulations and the full-Bayesian variable selection proposed by \cite{Linero18}. 

Section \ref{sec:est} introduces notation and a revision on tree-based Bayesian regression trees for causal inference. Section \ref{sec:analysis} specify how the propensity score and the Individual Conditional Expectation (ICE) Plots can be used to properly perform treatment effect analysis on Bayesian regression trees. Section \ref{sec:simul} have simulations on which the techniques introduced on the previous section are used. Finally, in Section \ref{sec:discuss} the results of the study are discussed, along with proposed extensions.

\section{Treatment Effect Estimation with Bayesian Regression Trees Models}
\label{sec:est}
Capital roman letters are used to denote random variables, while realizations are denoted in lower case. Vectors are denoted by a tilde on the top of the variable, and matrices are denoted by bold variables. Let $Y$ denote a scalar response, $Z$ denote a binary treatment effect and $X$ denote a vector of $p$ covariates $\left\{\tilde{X} = \left(X_1,...,X_p\right)\right\}$. In such way that the triplet $\left(Y_i,Z_i,\tilde{X}_i\right)$ denotes the observation of the $i$th individual of a sample size $n$.

Following the notation of \cite{Imbens04}, the $Y_i$ that has a $Z_i$ realization can be denoted by:

\begin{equation*}
	Y_i \equiv Y_i\left(Z_i\right) = 
	\begin{cases}
		Y_i\left(0\right) & \text{if $Z_i = 0$},\\
		Y_i\left(1\right) & \text{if $Z_i = 1$}.
	\end{cases}
\end{equation*}
\\
It is important to notice that only $Y_i(0)$ or $Y_i(1)$ can be observed, while the unobserved $Y_i\left(Z_i\right)$ is called counterfactual. 

In the paper \textit{strong ignorability} \citep{Rubin83} is assumed to hold. This condition require two assumptions:

\begin{assumption} - \text{Unconfoundedness} \label{as:1}
	\item $Y\left(0\right), Y\left(1\right) \Perp Z \mid \tilde{X}$
\end{assumption}

\begin{assumption} - \text{Overlap} \label{as:2}
	\item $0 < \mathbb{P}(Z=1 \mid \tilde{X}) < 1$
\end{assumption}

The first assumption guarantees that there are no unmeasured confounders in the analysis, while the second assumption assures that there is a positive probability of assigning each treatment to every individual in the population, always enabling the existence of the counterfactual, thus making it possible to estimate the treatment effect.

Using the framework adopted by \cite{Hahn17} for expressing treatment effects, it follows that the Individual Treatment Effect (ITE) can be represented as

\begin{equation}
		\alpha\left(\tilde{x}_i\right)=\mathrm{E}\left(Y_i \mid \tilde{x}_i,Z_i=1 \right)-\mathrm{E}\left(Y_i \mid \tilde{x}_i,Z_i=0 \right), \label{eq:ite}
\end{equation}
\\
\noindent
where $\alpha\left(\tilde{x}_i\right)$ is the ITE for the $i$th individual of the sample.

As in \cite{Hill11}, the Conditional Average Treatment Effect (CATE) is estimated by

\begin{equation}
\frac{1}{n}\sum_{i=1}^{n}\left[\mathrm{E}\left(Y_{i}\left(1\right) \mid \tilde{x}_{i}\right)-\mathrm{E}\left(Y_{i}\left(0\right) \mid \tilde{x}_{i} \right)\right], \label{eq:ate}
\end{equation}
\\
\noindent
which is equivalent to the average of the individual treatment effects of the sample.

\subsection{BART Model}
\label{subsec:bart}
\cite{CGM10} introduced the BART model, which is a Bayesian nonparametric ``sum-of-trees'' model, where each tree is a weak learner constrained by a regularization prior. The individual tree is formed by a set of binary splits $\left\{T\right\}$ from the set of covariates $\left\{\tilde{x} = \left(x_1,...,x_p\right)\right\}$, and a set terminal nodes $\left\{B = \left(\mu_1,...,\mu_b\right)\right\}$. Each split is of the form $\left\{x_l \leq c\right\}$ vs $\left\{x_l > c\right\}$ for continuous variables.

\begin{figure}
	\begin{center}
		\includegraphics[scale=0.65]{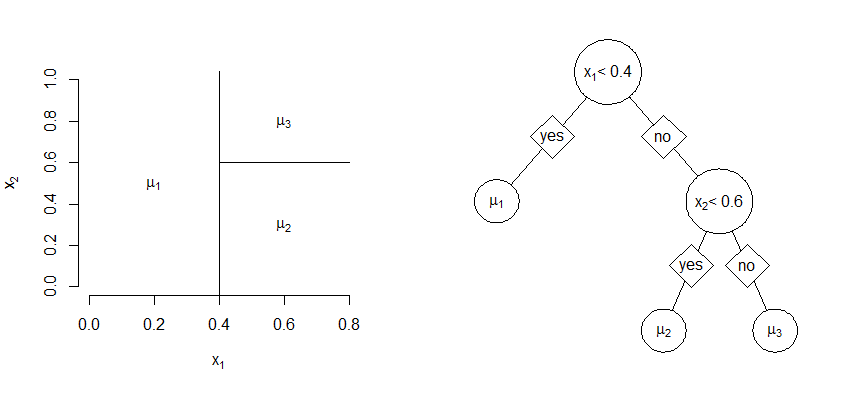}
	\end{center}
	\caption{(Left) A partitioned space. (Right) The tree with the corresponding binary splits to the partitioned space. \label{fig:extree}}
\end{figure}

The model can be expressed as

\begin{equation}
	Y_i = \sum_{j=1}^{m}{g\left(\tilde{x}_i;T_j;M_j\right)} + \varepsilon_i, \ \ \ \ \varepsilon_i\sim\mathcal{N}\left(0,\sigma^{2}\right), \label{eq:bart} 
\end{equation}

\noindent
where $m$ is the number of trees on the sum, and $g(\tilde{x}_i;T_j;M_j)$ is the tree function from \cite{CGM98}, which assigns a value $\left\{\mu_{kj} \in B_j\right\}$ from the $j$th tree to $\tilde{x}_i$ as seen in the example from Figure \ref{fig:extree}.

The priors of the model can be specified as

\begin{equation}
	p\left(\left(T_{1}, M_{1}\right), \left(T_{2}, M_{2}\right), ... , \left(T_{m}, M_{m}\right), \sigma \right) = \left[\prod_{j=1}^{m}{p\left(M_{j}\mid T_{j}\right)p\left(T_{j}\right)}\right]p\left(\sigma \right), \label{eq:prior}
\end{equation}
\noindent
and

\begin{equation}
	p\left(M_{j}\mid T_{j}\right) = \prod_{k=1}^{|M_{j}|}{p\left(\mu_{kj} \mid T_{j} \right)}, \label{eq:prior2}
\end{equation}
\\
\noindent
where $|M_{j}|$ is the cardinality of the set $M_{j}$.

The $p\left(\mu_{kj} \mid T_{j} \right)$ prior works as a regularization prior which constrains each tree to be a ``weak learner'', which is a model that performs better than chance. For convenience, $\tilde{y}$ is scaled between $-0.5$ and $0.5$, so that the prior given by

\begin{equation}
	\mu_{kj} \sim \mathcal{N}\left(0, \sigma_{\mu}^{2}\right), \ \ \ \ \text{with} \ \sigma_{\mu} = \frac{0.5}{k\sqrt{m}}, \label{eq:muprior}
\end{equation}
\\
\noindent
helds, for the default setting of $k=2$, $95\%$ probability that the expected value of the response lies within the interval $\left(-0.5,0.5\right)$. The hyperparameter $k$ can, also, be chosen by cross-validation.

The $p\left(\sigma \right)$ prior is given by

\begin{equation}
	\sigma^{2} \sim \frac{\nu\lambda}{\chi_{\nu}^{2}}, \label{eq:sigmaprior}
\end{equation}
\\
\noindent
which is an inverse chi-square distribution. The hyperparameter $\lambda$ is given in a way such that $\mathbb{P}\left(\sigma < \hat{\sigma}\right) = q$, where $\hat{\sigma}$ is an initial guess based on the data. \cite{CGM10} recommends the default setting as $\left(\nu,q\right)=\left(3,0.90\right)$.

The $p\left(T_{j}\right)$ prior has three parts. The first one is the probability of a split in a determined node, which is given by

\begin{equation}
	\eta\left(1+d\right)^{-\beta}, \ \ \ \ \eta \in \left(0,1\right), \ \beta \in [0, \infty), \label{eq:tprior}
\end{equation}
\\
\noindent
where $\eta$ and $\beta$ are hyperparameters with default values $\eta = 0.95$ and $\beta = 2$, and $d$ is the depth of tree $\left\{d \in \left(0,1,...\right)\right\}$. The second part, which is the probability of selecting a determined variable to perform a split, is given by a discrete Uniform hyperprior. Finally, the third part is given by a discrete Uniform distribution over the possible splits of the variable chosen for the split. 

The posterior of the model can be expressed as

\begin{equation}
	p\left(\left(T_1,M_1\right),\left(T_2,M_2\right),...,\left(T_m,M_m\right), \sigma \mid \tilde{y}, \textbf{x} \right). \label{eq:bartpost}
\end{equation}

One way to sample from this posterior distribution is through the Bayesian backfitting algorithm introduced by \cite{Hastie00}, which, basically, is a Gibbs sampler drawing $\left(T_j,M_j\right)$ with $j \in \{1,2,...m\}$, conditionally on $\left(\left(T_{(j)},M_{(j)}\right), \sigma \right)$, where $\left(T_{(j)},M_{(j)}\right)$ is the set of $m-1$ trees with its associated terminal node parameters except $\left(T_j,M_j\right)$.

To perform a draw from $p\left(\left(T_{j}, M_{j}\right) \mid \left(T_{(j)},M_{(j)}\right), \sigma \right)$ it is important to notice that $\left(T_{(j)},M_{(j)}\right)$ have impact on $\left(T_{j},M_{j}\right)$ only through

\begin{equation}
	\tilde{R}_j \equiv \tilde{y} - \sum_{h \neq j}{g\left(\textbf{x};T_h;M_h\right)}, \label{eq:resid}
\end{equation}
\\
\noindent
in such a way that $p\left(\left(T_j,M_j\right) \mid \tilde{R}_j, \sigma \right)$ can be sampled using the framework of \cite{CGM98} for drawing samples of a single tree.

The authors introduced four proposals from which the trees can mix by using the Metropolis-Hastings algorithm: GROW (grow a split from a terminal node), PRUNE (collapse a split above two terminal nodes), CHANGE (change a rule from a nonterminal node) and SWAP (swap the rules of a parent and child nonterminal nodes). \cite{Pratola16} also presents the ROTATION and the PERTURB proposals as an alternative to improve the tree mixing. For further details of proposals GROW, PRUNE and CHANGE, see \cite{Bleich16}. 

\cite{Hill11} noted that as the BART captures interactions and nonlinearities with ease, handle a large number of covariates that are potential confounders, and have a stable default setup for its priors. So the model is a tool that can be applied to the causal inference setting, specially at the estimation of CATE, since the model estimates of ITE have shown great uncertainty.

\subsection{BCF Model}
\label{subsec:bcf}
One possible extension of the BART model can be achieved by adding linear components to the model \cite[p.~295]{CGM10}. The Bayesian Causal Forests (BCF) model introduced by \cite{Hahn17} follows this idea by using a linear combination of two BART models to estimate the value of the response $Y_i$.

The BCF model is given by

\begin{equation}
	Y_i = m\left(\tilde{x}_i,\hat{\pi}\left(\tilde{x}_i\right)\right) + \alpha\left(\tilde{x}_i,\hat{\pi}\left(\tilde{x}_i\right)\right)z_i + \varepsilon_i, \ \ \ \ \varepsilon_i\sim\mathcal{N}\left(0,\sigma^{2}\right), \label{eq:bcf}
\end{equation}
\\
\noindent
where $\alpha$ and $m$ are independent BART priors, $\tilde{x}_i$ is vector of $p$ covariates for the $i$th individual of the sample, and $\hat{\pi}\left(\tilde{x}_i\right)$ is the estimate of the propensity score $\pi\left(\tilde{x}_i\right)=\mathbb{P}\left(Z_i=1 \mid \tilde{x}_i\right)$ for the $i$th individual. The propensity score estimate for each individual is introduced in the model as a covariate with the main objective of reducing bias into the estimate of treatment effects. Its role is further explored at Section \ref{subsec:pscore}.

The function $m$ estimate the prognostic effect of each individual, while the function $\alpha$ is used to capture its treatment effect. The BCF has been designed this way because in the original BART setting there was no control on how the model varies in $Z$, and in this new reparametrization the $Z$ works like an indicator function for the $\alpha$, enabling the model to aggregate all the covariates interactions regarding treatment effects in the same function.

Since the functions $m$ and $\alpha$ are independent, each can have different priors that are adaptable according to its characteristics. Both functions held reasonable results with the default prior settings, but \cite{Hahn17} have chosen to do two main modifications regarding $\alpha$. The first modification has been made to support homogeneous treatment effects in the model by two amendments in the $T_j$ prior: Considering that the homogeneous treatment effects are represented by trees that are root nodes, the parameter $\eta$ is adapted to the probability of homogeneous effects $\left\{\alpha_0 \in \left(0,1\right)\right\}$ by solving $\left\{\alpha_0 = \left(1-\eta\right)^{L_\alpha}\right\}$, where $L_\alpha$ is the number of trees in the function $\alpha$; Setting $\beta = 3$ (instead of $\beta = 2$) to lower the split probability, since the $z_i$ works as an indicator function which can be compared to a first tree split in the variable $Z$, meaning that all trees in $\alpha$ actually start with depth of $1$. The second modification has been made through the implementation of a half-Cauchy hyperprior $\left\{\nu_{\alpha} \sim \mathcal{C}\left(0,\nu_0\right)_{+}\right\}$ to the scale parameter of $\alpha$, where $\alpha\left(\tilde{x}_{i}\right) \sim \mathcal{N}\left(0,\nu_{\alpha}\right)$. The default BART uses a constant to its scale parameter, but the change grants a way of avoiding spurious inferences. For further details on the half-Cauchy hyperprior, see \cite{Gelman06}.

Furthermore, the $\alpha$ posterior estimates can be used to analyze the treatment effects at individual level in a way that is possible to identify groups which have positive (or negative) treatment effects within a certain credible interval, allowing a kind of study that was not recommended in \cite{Hill11} framework due to the lack of estimates robustness.

\section{Treatment Effect Analysis}
\label{sec:analysis}
The BART model presented in Section \ref{subsec:bart} can be used to estimate the ITE by

\begin{equation}
	\alpha\left(\tilde{x}_{i}\right) = \mathrm{E}\left(Y_i \mid \tilde{x}_i,Z_i=1 \right)-\mathrm{E}\left(Y_i \mid \tilde{x}_i,Z_i=0 \right) = \hat{f}\left(\tilde{x}_i,1\right) - \hat{f}\left(\tilde{x}_i,0\right), \label{eq:treat2}
\end{equation}
\\
\noindent
where $\hat{f}\left(\tilde{x}_i,z_i\right)$ consist of the posterior draws from the prediction of the estimated model. 

In the case of the BCF model from Section \ref{subsec:bcf}, as the $\alpha$ function is estimated separately from the prognostic effect, the BART prior of $\alpha$ already gives the posterior draws from the estimated model as an output. It should be noted that since both models outputs are given by posterior draws, it is straightforward to construct credible intervals for the estimated treatment effects with the use of quantiles from these draws.

The following subsections introduce some methods and tools to assist the analysis of treatment effects in these models.

\subsection{The RIC and the Role of the Propensity Score}
\label{subsec:pscore}

The term  ``regularization-induced confounding'' (RIC) was introduced by \cite{Hahn18} into the setting of linear models with homogeneous treatment effects and further expanded by \cite{Hahn17} to the BART models, which despite having a good predictive performance, have shown biased treatment effect estimation and lack of robustness at the individual level estimates when applied to the causal settings. \cite{Hahn17} tries to avoid the RIC phenomenon by including the estimate of the propensity score as a covariate in the BART and the BCF models.

The propensity score was first introduced by \cite{Rubin83} as an assisting tool to reduce bias in observational studies by balancing the data according to the probability of assigning a treatment to an individual given its vector of covariates. The authors also introduced matched sampling, subclassification, and covariance adjustment techniques by using the propensity score. In other words, the propensity score was created to deal with the problem of treatment effects bias in observational studies.

As shown by \cite{Hahn17}, the inclusion of the propensity score allows the tree-based models to adapt more easily in cases where the data exhibits complex confounding. Since the propensity score naturally simplifies the number of required splits in a context of parsimonious trees, it allows the model to focus on other interactions between the variables, reducing bias and improving the predictions of the model. 

Under the sparsity setting it is possible to assessthe use of the propensity score in the model by the variable selection framework introduced by \cite{Linero18} in its Dirichlet Additive Regression Trees (DART) models, which assigns a sparsity-inducing Dirichlet hyperprior (instead of an Uniform hyperprior) on the probability of choosing a variable to split on, which is given by

\begin{equation*}
(s_1,...s_{P})\sim\mathcal{D}\left(\frac{\theta}{P}, \frac{\theta}{P}, ..., \frac{\theta}{P} \right),
\end{equation*}

where $P$ is the number of covariates in the model and $\theta$ is given by the prior

\begin{equation*}
\frac{\theta}{\theta + \rho} \sim Beta\left(a,b \right),
\end{equation*}

with $a=0.5$, $b=1$ and $\rho=P$ in this paper.

The author suggests two approaches to perform the variable selection: The first one is to look into the posterior draws from the Dirichlet hyperprior and the second one is to use the method of variable selection proposed by \cite{Berger04} and calculate the posterior inclusion probability ($PIP$) for each variable.

The former approach is straightforward, since the posterior draws allow the construction of credible intervals for the probability of choosing any variable. The latter approach can be performed by simple verifying, for each iteration of the MCMC, if the variable $l$ is used in at least one splitting rule of the tree ensemble. If it is used, the value $1$ is assigned, and, if not, $0$ is assigned. The $PIP_l$ is the mean of these indicator functions for variable $l$ over all iterations of the MCMC, and the variable $l$ will be selected if $PIP_l > 0.5$.

\subsection{A Visualization Tool: ICE Plots}
\label{subsec:ice}

\cite{Friedman01} introduced the Partial Dependence Plot (PDP) in order to allow the visualization of the impact that a variable have in the response, performing an sensitivity analysis. The main reason for the development of such a tool is due to the difficulty to interpret parameters in machine learning methods, more specifically, the models known as ``black box'' models, such as Support Vector Machines, Random Forests, Neural Networks, etc.

The partial dependence function for the $i$th observation is defined as

\begin{equation}
	\hat{f}\left(x_{Si}\right)=\frac{1}{n} \sum_{j = 1}^{n} \hat{f}\left(x_{Si}, \tilde{x}_{Cj}\right), \label{eq:pdp}
\end{equation}
\\
\noindent
where $n$ is the sample size, $\hat{f}(.)$ is the prediction function from the ``black box'' model, $S \in \left\{1,...,p\right\}$ is the index of one of variables of interest, $C = S^{\mathsf{c}}$, $x_{Si}$ is a scalar from the variable of interest from the $i$th observation of the training data, and $\tilde{x}_{Cj}$ is the vector of covariates from the subset $C$ of the observation $j$ from the training data. The curve is made from the partial dependence functions of all observations from $\tilde{x}_{S}$. In general, $S \subset \left\{1,...,p\right\}$ with $|S|\geq 1$, but there are no means of plotting the PDP for $|S|>2$.

\cite{CGM10} suggest the use of PDP to analyze the marginal effect of the variables in relation to the response. It must be noted that the implementation of this technique in BART models actually give draws from the posterior distribution, thus, it is easy to acquire the credible intervals for the PDP curve.
 
This tools have already been applied to BART models in the treatment effect setting in \cite{Green12}, where the authors create curves regarding each variable in relation to the CATE, which takes the place of the response variable. This is possible due to a slight modification in the algorithm by using

\begin{equation}
\hat{f}_{CATE}\left(x_{Si}\right)=\frac{1}{n} \sum_{j = 1}^{n} \left[ \hat{f}\left(x_{Si}, \tilde{x}_{Cj}, z_i = 1\right)  - \hat{f}\left(x_{Si}, \tilde{x}_{Cj}, z_i = 0\right)  \right]. \label{eq:pdpcate}
\end{equation}

But since this tool only evaluates the mean of the response, it cannot be used into the analysis of individual treatment effects.

\cite{Goldstein15} introduced the ICE Plots by noting that since the PDP curve could conceal heterogeneous effects in the response variable, it was necessary a tool for analyzing the marginal effect at each individual. By using the same notation of equation \eqref{eq:pdp} the ICE function for the $i$th individual is defined by

\begin{equation}
\hat{f}\left(x_{Sij}\right)=\hat{f}\left(x_{Sj}, \tilde{x}_{Ci}\right), \label{eq:ice}
\end{equation}
\\
\noindent
where the ICE curve for the $i$th individual is formed by calculating the ICE function for every $j$ of the sample. This way, there will be $n$ curves in the plot, and the PDP curve can be obtained by averaging the ICE curves.

This analysis can be very useful in the context of treatment effects, since in the setting of heterogeneous treatment effects it is possible to look for indications of groups that have different reactions to the applied treatment. Like in equation \eqref{eq:pdpcate}, the ICE curves can be adapted to the ITE setting by estimating

\begin{equation}
\hat{f}_{ITE}\left(x_{Sij}\right)= \left[ \hat{f}\left(x_{Sj}, \tilde{x}_{Ci}, z_i = 1\right)  - \hat{f}\left(x_{Sj}, \tilde{x}_{Ci}, z_i = 0\right)  \right]. \label{eq:icecate}
\end{equation}
\\
\noindent
instead of using the formula described in equation \eqref{eq:ice}.

Examples from the ICE Plots and the PDPs can be found on Section \ref{subsec:example}.

\section{Simulations}
\label{sec:simul}
In order to assess the tools that have been proposed in Section \ref{sec:analysis}, some simulations were performed. In Section \ref{subsec:example} the simulations were made in order to corroborate the use of the propensity score as a covariate in BART models by using the ICE Plots to evaluate the ITE of the sample. In Section \ref{subsec:high} the  simulations were performed in the sparsity setting, to evaluate how often the propensity score is used by the model in relation to the other variables. Only the BART models were studied at Sections \ref{subsec:example} and \ref{subsec:high}, but these methods can be extended to the BCF models as well. Furthermore, in Section \ref{subsec:simasses} the CATE and ITE estimates of BART and BCF models were evaluated. 

\cite{Zigler14} pointed that since the propensity score carries uncertainty about its estimations, so it is natural to advocate the use of a Bayesian framework in it. Following \cite{Hahn17}, the probit-BART posterior mean is used as the propensity score estimate for each individual in the simulations. As a comparative, the frequentist approach of estimating the propensity score by Generalized Linear Models with \textit{logit} link was used.

All calculations were performed in R version 3.4.3 \citep{R} by using the \textit{packages} BART \citep{BART}, bcf \citep{bcf} and ICEbox \citep{Goldstein15}. As in \cite{Hill11}, the priors, hyperparameters, and hyperpriors were held under default setting in this paper, but cross-validation can be performed in order to improve results.

\subsection{Simulation Based on Real Data}
\label{subsec:example}

In this scenario, the following models were analyzed:

\begin{itemize}
	\item Vanilla: $Y_i$ estimated by a BART model using $\tilde{x}_i$ as covariates;
	\item Oracle (Oracle-BCF): $Y_i$ estimated by a BART (BCF) model using $\tilde{x}_i$ and $\pi\left(\tilde{x}_i\right)$, the true value of the propensity score, as covariates;
	\item PS-BART (PS-BCF): $Y_i$ estimated by a BART (BCF) model using $\tilde{x}_i$ and $\hat{\pi}\left(\tilde{x}_i\right)$, estimated by the posterior mean of the probit-BART, as covariates;
	\item GLM-BART (GLM-BCF): $Y_i$ estimated by a BART (BCF) model using $\tilde{x}_i$ and $\hat{\pi}\left(\tilde{x}_i\right)$, estimated by GLM (which is considered as a naive approach to estimate the propensity score), as covariates;
	\item Rand-BART (Rand-BCF): $Y_i$ estimated by a BART (BCF) model using $\tilde{x}_i$ and $\hat{\pi}\left(\tilde{x}_i\right)$, given by a random Uniform distribution, as covariates.
\end{itemize}

This simulation is based on \cite{Hill11}, and uses the Infant Health and Development Program (IHDP) dataset ($n=985$). 
In order to simplify the simulation, only the continuous covariates are used: birth weight ($x_1$); head circunference ($x_2$); weeks born preterm ($x_3$); birth order ($x_4$); neonatal health index ($x_5$) and age of the mother ($x_6$). 
The response surface was generated by

\begin{equation*}
Y_{i} = \beta_{1}x_{i1} + ... + \beta_{6}x_{i6} + \mu_{i} + Z_{i}\alpha_{i} + \epsilon_{i}, \ \ \ \ \epsilon_{i}{\sim}\mathcal{N}(0,0.5^{2}).
\end{equation*}

The predictors were standardized for data generation, and the $\beta_i$'s were sampled from (0, 1, 2, 3, 4) with probabilities (0.05, 0.1, 0.15, 0.2, 0.5).

The true propensity score and the real treatment effects were generated based on \cite{Hahn17} example as it follows,

\begin{equation*}
	\mu_{i}=\textbf{1}(x_{i1}<x_{i2})-\textbf{1}(x_{i1} \geq x_{i2}),
\end{equation*}
\begin{equation*}
	P(Z_{i}=1 \mid x_{i1},...,x_{i6}) = \Phi(\mu_{i}),
\end{equation*}
\begin{equation*}
	\alpha_{i} = 0.5*\textbf{1}(x_{i3} > -3/4) + 0.25*\textbf{1}(x_{i3} > 0) + 0.25*\textbf{1}(x_{i3}>3/4).
\end{equation*}

This simulation was replicated $1000$ times in order to assessthe results. In each model the first $1000$ draws from the MCMC were treated as burn in, while the posterior draws had size $2000$.

\begin{figure}
	\begin{center}
		\includegraphics[scale=0.5]{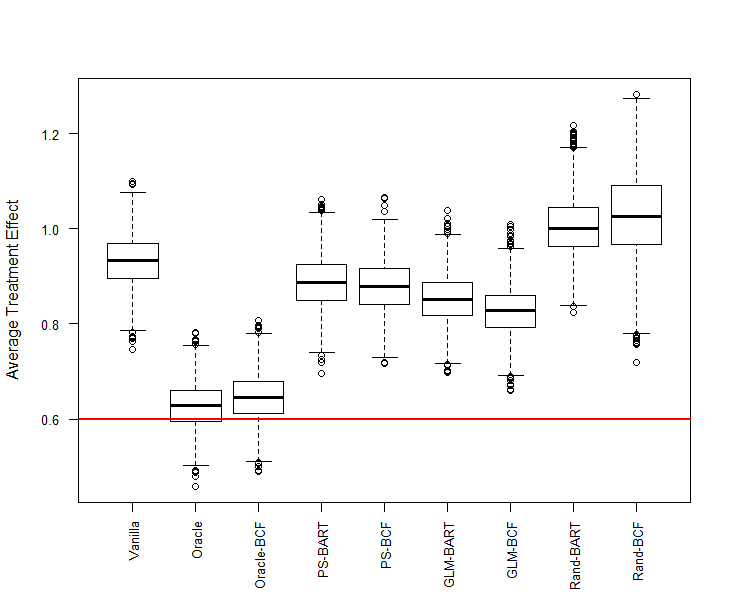}
	\end{center}
	\caption{Hill example - Boxplots of the posterior CATE for each model from one iteration. The red line represents the true CATE. \label{fig:ate}}
\end{figure}

The Figure \ref{fig:ate} is composed by the boxplots of posterior CATE for each model for the one replication of the simulation, but it is important to point out that similar results were found for the other replications. The red line is the real CATE for this specific iteration. 

In general, the BART and the BCF models held similar results. The Vanilla model posterior CATE estimates apparently are impacted by the RIC phenomenon, so the model performed poorly. The Oracle models, as expected, had a good performance due to the inclusion of the propensity score as a covariate. The PS and the GLM models performed slightly better than the Vanilla model, indicating that the inclusion of the propensity score had a positive impact on the model, but the uncertainty associated with the estimation of propensity score contributed negatively on the CATE estimates. The Rand models held the worst results in the simulation due to the inclusion of an irrelevant variable and the lack of a propensity score estimate among its covariates.

\begin{figure}
	\begin{center}
		\includegraphics[scale=0.32]{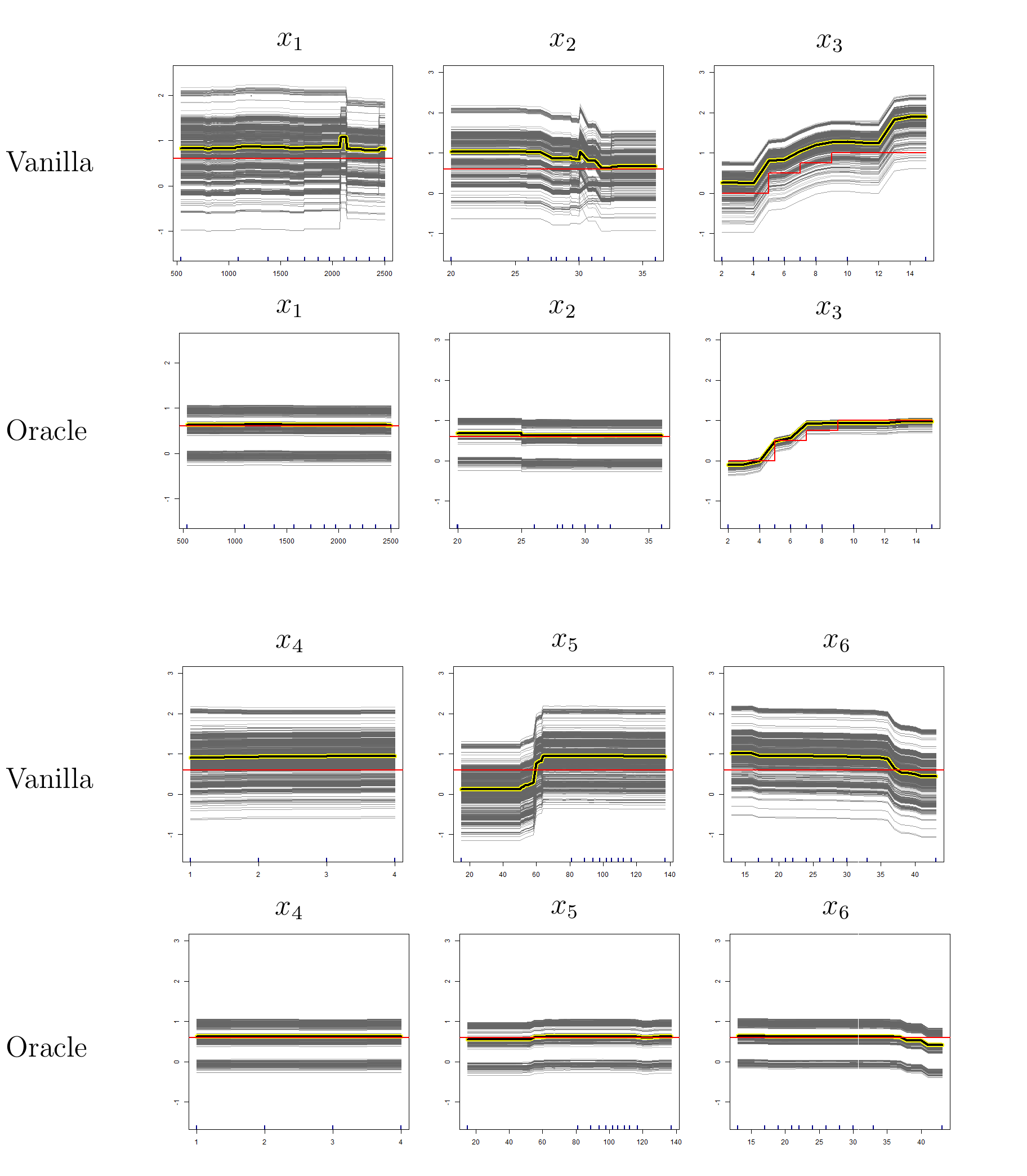}
	\end{center}
	\caption{Hill Example - ICE Plots of Vanilla and Oracle models for the impact of each variable on the estimated Treatment Effect. The red line represents the true average treatment effect. The black line with yellow borders is the PDP curve. The gray lines are the individual ICE curves. \label{fig:icevanillaoracle}}
\end{figure}

As seen in the ICE Plots of the Vanilla and Oracle models from Figure \ref{fig:icevanillaoracle}, the inclusion of the true propensity score as a covariate in greatly reduces the uncertainty over the individual treatment effects, allowing the visualization of different groups of individuals and eliminating most of spurious effects from the other covariates.

\begin{figure}
	\begin{center}
		\includegraphics[scale=0.5]{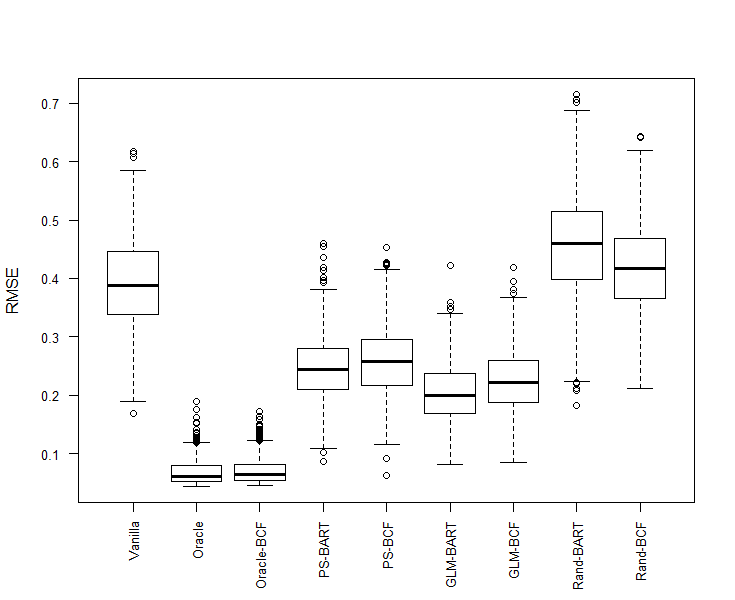}
	\end{center}
	\caption{Hill example - Boxplots of the CATE RMSE for each model calculated over 1000 simulations. \label{fig:atermse}}
\end{figure}

\begin{figure}
	\begin{center}
		\includegraphics[scale=0.5]{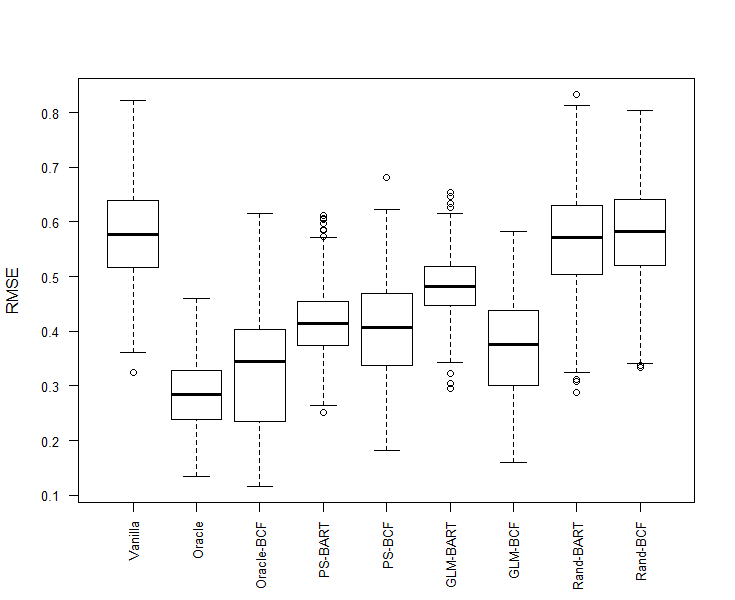}
	\end{center}
	\caption{Hill example - Boxplots of the ITE RMSE for each model calculated over 1000 simulations. \label{fig:itermse}}
\end{figure}

Furthermore, across $1000$ replications, the use of the propensity score as a covariate is corroborated, since the inclusion of true (estimated) propensity score greatly (slightly) improved the performance of the model, as seen in the CATE and ITE RMSE boxplots in Figures \ref{fig:atermse} and \ref{fig:itermse}. The means and the standard deviations for average RMSE of these models over the replications can be found on Table \ref{table:hill}.

\begin{table}
	\centering
	\begin{tabular}{ |c|c|c|c|c| }
		\hline
		\multirow{2}{*}{Model} & \multicolumn{2}{|c|}{BART} & \multicolumn{2}{|c|}{BCF} \\
		\cline{2-5}
		& CATE RMSE & ITE RMSE & CATE RMSE & ITE RMSE \\ \hline
		\hline
		\multirow{2}{*}{Vanilla} & 0.391 & 0.577 & - & - \\
		& (0.076) & (0.086) & - & - \\ \cline{1-5}
		\multirow{2}{*}{Oracle} & 0.068 & 0.284 & 0.071 & 0.322 \\
		& (0.021) & (0.063) & (0.021) & (0.103) \\ \cline{1-5}
		\multirow{2}{*}{PS} & 0.246 & 0.414 & 0.258 & 0.403 \\
		& (0.052) & (0.059) & (0.057) & (0.086) \\ \cline{1-5} 
		\multirow{2}{*}{GLM} & 0.203 & 0.482 & 0.224 & 0.370 \\
		& (0.049) & (0.053) & (0.054) & (0.085) \\ \cline{1-5}
		\multirow{2}{*}{Rand} & 0.416 & 0.580 & 0.453 & 0.566 \\
		& (0.074) & (0.085) & (0.090) & (0.093) \\ \hline
		
	\end{tabular}
	\caption{Hill example - Model assessment through the means of CATE RMSE, and ITE RMSE over replications. Standard deviation is given in parenthesis.}
	\label{table:hill}
\end{table}

Also, the BART and the BCF models held similar results across simulations, but for the ITE RMSE, the BCF model seems to have a better performance under the uncertainty of the estimation of the propensity score, despite having a higher variance across the RMSE calculated over the replications. Nevertheless, the BART model was superior in the Oracle scenario.

\subsection{Sparse Data Examples}
\label{subsec:high}

Under the sparse setting, the following models were analyzed:

\begin{itemize}
	\item Vanilla (Vanilla-DART): $Y_i$ estimated by a BART (DART) model using $\tilde{x}_i$ as covariates;
	\item Oracle (Oracle-DART): $Y_i$ estimated by a BART (DART) model using $\tilde{x}_i$ and $\pi\left(\tilde{x}_i\right)$, the true value of the propensity score, as covariates;
	\item PS-BART (PS-DART): $Y_i$ estimated by a BART (DART) model using $\tilde{x}_i$ and $\hat{\pi}\left(\tilde{x}_i\right)$, estimated by the posterior mean of the probit-BART (probit-DART), as covariates;
	\item GLM-BART (GLM-DART): $Y_i$ estimated by a BART (DART) model using $\tilde{x}_i$ and $\hat{\pi}\left(\tilde{x}_i\right)$, estimated by GLM, as covariates;
	\item Rand-BART (Rand-DART): $Y_i$ estimated by a BART (DART) model using $\tilde{x}_i$ and $\hat{\pi}\left(\tilde{x}_i\right)$, given by a random Uniform distribution, as covariates;
\end{itemize}

In order to acknowledge the propensity score role in the model, a method of variable selection was performed. Selected variables were those whose presented $PIP > 0.5$. To assessthe performance of the variable selection, following \cite{Linero18} and \cite{Bleich14}, Precision, Recall, and $F_{1}$ were used. These measures are defined by,

\begin{equation*}
Precision = \frac{TP}{TP+FP}, \ \ \ \ Recall = \frac{TP}{TP+FN}, \ \ \ \ F_{1}=2 \times \frac{Precision \times Recall}{Precision + Recall},
\end{equation*}
\\
\noindent
where $TP$ is True Positive, $FP$ is False Positive, and $FN$ is False Negative.

These simulations were replicated $100$ times in order to assessthe results. In each model the first $5000$ draws from the MCMC were treated as burn in, while the posterior draws had size $1000$, with thinning being set to $50$. For all simulations, $n=1000$.

\subsubsection{Hahn Simulation under Sparsity}
\label{subsubsec:hahnsparse}

The example based on \cite{Hahn17} simulation was generated as it follows,
\begin{equation*}
Y_{i} = 0.1x_{i1} + 0.1x_{i2} + \mu_{i} + Z_{i}\alpha_{i} + \epsilon_{i}, \ \ \ \ \epsilon_{i}\sim \mathcal{N}(0,\sigma^2),
\end{equation*}
\begin{equation*}
x_{i1},x_{i2},...,x_{i98}\sim \mathcal{N}(0,1),
\end{equation*}
\begin{equation*}
\mu_{i}=\textbf{1}(x_{i1}<x_{i2})-\textbf{1}(x_{i1} \geq x_{i2}),
\end{equation*}
\begin{equation*}
P(Z_{i}=1 \mid x_{i1},x_{i2}) = \Phi(\mu_{i}),
\end{equation*}
\begin{equation*}
\alpha_{i} = 0.5*\textbf{1}(x_{i3} > -3/4) + 0.25*\textbf{1}(x_{i3} > 0) + 0.25*\textbf{1}(x_{i3}>3/4),
\end{equation*}
\begin{equation*}
\sigma = \frac{\theta^{(n)} - \theta^{(1)}}{8}, \ \ \ \ \theta_{i} = \mu_{i} + \alpha_{i}\Phi\left(\mu_{i}\right),
\end{equation*}
\\
\noindent
where $\theta^{(n)} = \max\left(\theta_1,...,\theta_n\right)$ and $\theta^{(1)} = \min\left(\theta_1,...,\theta_n\right)$.

\begin{figure}
	\begin{center}
		\includegraphics[scale=0.5]{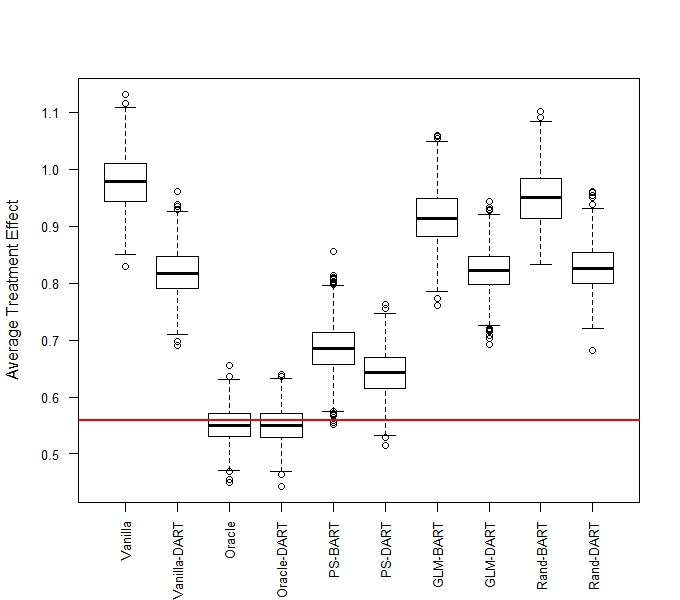}
	\end{center}
	\caption{Hahn example - Boxplots of the posterior CATE for each model from one iteration. The red line represents the true CATE. \label{fig:dart1}}
\end{figure}

Like in the previous example, the Figure \ref{fig:dart1} is composed by the boxplots of posterior CATE for each model for the one replication of the simulation. The red line is the real CATE for this specific iteration. 

Under the sparse setting, the DART models, in general, had a better performance at estimating the CATE. The Vanilla model estimates remained close to the Rand models estimates. The GLM models performed slightly better than the Vanilla model, but the inclusion of many irrelevant variables on the model have shown a negative impact on the estimates. The PS models have performed significantly better than the GLM models, while, as expected, the Oracle models have shown the best results.

\begin{figure}
	\begin{center}
		\includegraphics[scale=0.5]{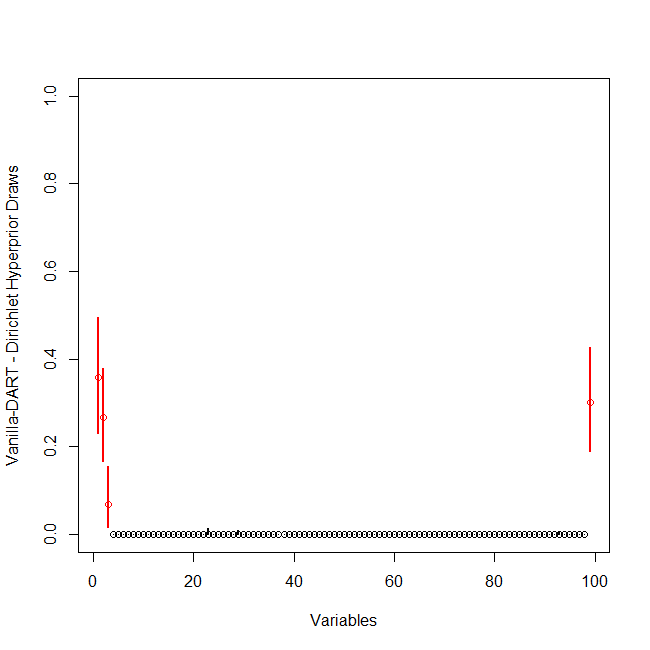}
	\end{center}
	\caption{Hahn example - Posterior draws from Vanilla-DART Dirichlet hyperprior with 95\% credible intervals.  In red: $x_1$, $x_2$, $x_3$, and $z$, respectively. \label{fig:vs0}}
\end{figure}

\begin{figure}
	\begin{center}
		\includegraphics[scale=0.5]{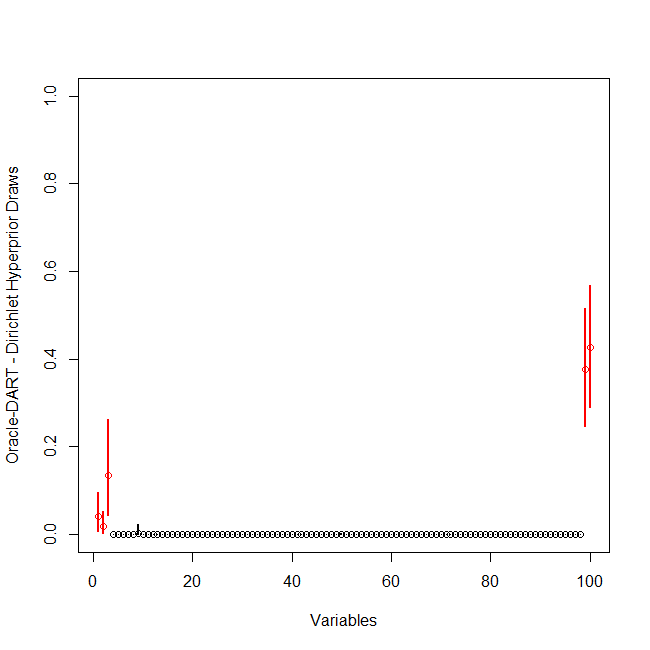}
	\end{center}
	\caption{Hahn example - Posterior draws from Oracle-DART Dirichlet hyperprior with 95\% credible intervals.  In red: $x_1$, $x_2$, $x_3$, $\pi(\tilde{x})$ and $z$, respectively. \label{fig:vs1}}
\end{figure}

\begin{figure}
	\begin{center}
		\includegraphics[scale=0.5]{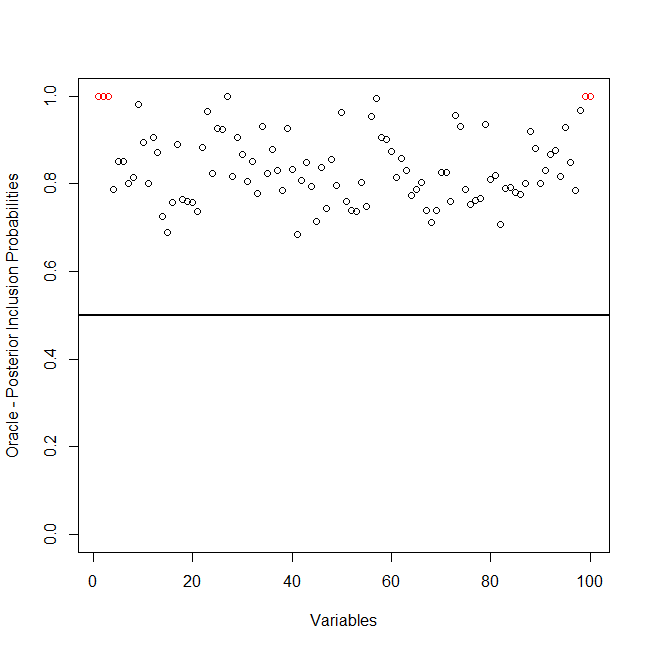}
	\end{center}
	\caption{Hahn example - Posterior Inclusion Probability of Oracle model. In red: $x_1$, $x_2$, $x_3$, $\pi(\tilde{x})$, and $z$, respectively.  \label{fig:vs2}}
\end{figure}

\begin{figure}
	\begin{center}
		\includegraphics[scale=0.5]{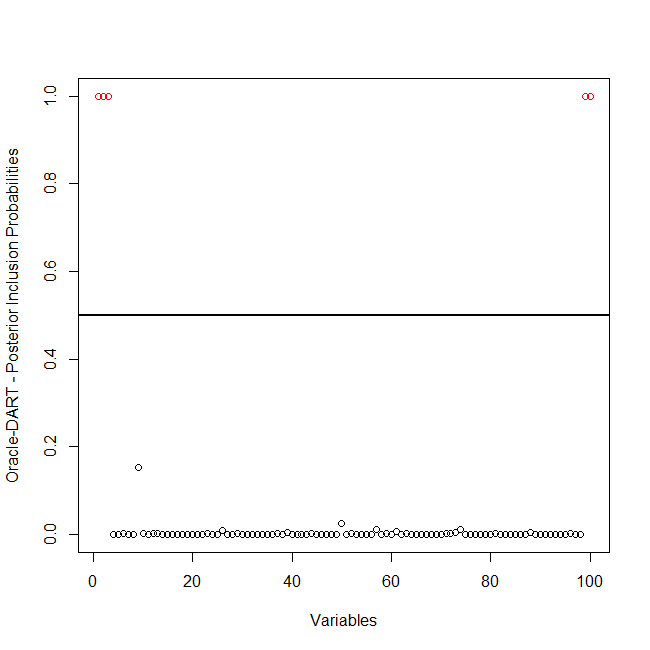}
	\end{center}
	\caption{Hahn example - Posterior Inclusion Probability of Oracle-DART model. In red: $x_1$, $x_2$, $x_3$, $\pi(\tilde{x})$, and $z$, respectively.  \label{fig:vs3}}
\end{figure}

Figures \ref{fig:vs0} and \ref{fig:vs1} represents the Dirichlet hyperprior draws for the Vanilla-DART and Oracle-DART models. The inclusion of the propensity score in the Oracle model allows the tree ensemble to focus on the variable $x_3$, which determines the treatment effect of each individual, instead of trying to figure out how the variables $x_1$ and $x_2$ are related. Figures \ref{fig:vs2} and \ref{fig:vs3} represents BART and DART $PIP$ estimation from the Oracle models. While BART tends to use all the variables in the tree ensemble, the Dirichlet hyperprior on DART select only a few among the possible covariates.

\begin{table}
	\centering
	\begin{tabular}{ |c|c|c|c|c|c|c|c| }
		\hline
		& Model & CATE RMSE & ITE RMSE & Precision & Recall & $F_1$ & PS-Usage \\ \hline
		\hline
		\multirow{12}{*}{BART} 
		& \multirow{2}{*}{Vanilla} & 0.466 & 0.413 & 0.040 & 1.000 & 0.078 & - \\
		& & (0.062) & (0.161) & (0.000) & (0.000) & (0.000) & - \\ \cline{2-8}
		& \multirow{2}{*}{Oracle} & 0.046 & 0.346 & 0.050 & 1.000 & 0.095 & 100.000 \\
		& & (0.017) & (0.176) & (0.000) & (0.000) & (0.000) & (0.000) \\ \cline{2-8}
		& \multirow{2}{*}{PS} & 0.235 & 0.339 & 0.050 & 1.000 & 0.095 & 100.000 \\
		& & (0.063) & (0.103) & (0.000) & (0.000) & (0.000) & (0.000) \\ \cline{2-8} 
		& \multirow{2}{*}{GLM} & 0.397 & 0.378 & 0.050 & 1.000 & 0.095 & 100.000 \\
		& & (0.057) & (0.123) & (0.000) & (0.000) & (0.000) & (0.000) \\ \cline{2-8}
		& \multirow{2}{*}{Rand} & 0.464 & 0.431 & 0.050 & 1.000 & 0.095 & 88.418 \\
		& & (0.062) & (0.172) & (0.000) & (0.000) & (0.000) & (0.054) \\ \cline{2-8}
		& \multirow{2}{*}{Probit} & - & - & 0.158 & 1.000 & 0.271 & - \\
		& & - & - & (0.034) & (0.000) & (0.049) & - \\ \hline
		\hline
		\multirow{12}{*}{DART} 
		& \multirow{2}{*}{Vanilla} & 0.326 & 0.437 & 0.991 & 1.000 & 0.995 & - \\
		& & (0.052) & (0.200) & (0.047) & (0.000) & (0.027) & - \\ \cline{2-8}
		& \multirow{2}{*}{Oracle} & 0.045 & 0.361 & 0.998 & 0.986 & 0.991 & 100.000 \\
		& & (0.016) & (0.185) & (0.016) & (0.051) & (0.030) & (0.000) \\ \cline{2-8}
		& \multirow{2}{*}{PS} & 0.127 & 0.385 & 0.991 & 0.852 & 0.913 & 100.000 \\
		& & (0.043) & (0.184) & (0.041) & (0.088) & (0.051) & (0.000) \\ \cline{2-8} 
		& \multirow{2}{*}{GLM} & 0.303 & 0.418 & 0.997 & 0.922 & 0.955 & 62.935 \\
		& & (0.056) & (0.167) & (0.023) & (0.098) & (0.055) & (0.444) \\ \cline{2-8}
		& \multirow{2}{*}{Rand} & 0.326 & 0.441 & 0.995 & 0.800 & 0.886 & 0.262 \\
		& & (0.053) & (0.201) & (0.038) & (0.000) & (0.018) & (0.009) \\ \cline{2-8}
		& \multirow{2}{*}{Probit} & - & - & 0.997 & 1.000 & 0.998 & - \\
		& & - & - & (0.033) & (0.000) & (0.020) & - \\ \hline
		\hline
		
	\end{tabular}
	\caption{Hahn example - Model assessment through the means of CATE RMSE, ITE RMSE, Precision, Recall, $F_1$, and usage of the propensity score over replications. Standard deviation is given in parenthesis.}
	\label{table:hahn}
\end{table}

The performance of the models over the replications is evaluated at Table \ref{table:hahn}, along with the Probit model used to estimate the propensity score. Variables were selected via $PIP$. For Precision, Recall and $F_1$, the value $1.0$ indicates a perfect adjustment. PS-Usage indicates the mean of the proportion of times that the propensity score estimation was used in the model, and if the model had an estimation of the propensity score, regardless of misspecification, the estimation was considered a relevant variable in this analysis. All measurements are given by the mean over the replications, with standard deviation in parentheses.

\subsubsection{Friedman Function under Sparsity}
\label{subsubsec:friedmanhigh}

The simulation adapted from \cite{Friedman91} example was generated as it follows,
\begin{equation*}
Y_{i} = 10\sin(\pi x_{i1} x_{i2}) + 20(x_{i3}-0.5)^{2} + 10x_{i4} + 5x_{i5} + \mu_{i} + Z_{i}\alpha_{i} + \epsilon_{i}, \ \ \ \ \epsilon_{i}\sim \mathcal{N}(0,\sigma^2),
\end{equation*}
\begin{equation*}
x_{i1},x_{i2},...,x_{i98}\sim \mathcal{U}(0,1),
\end{equation*}
\begin{equation*}
\mu_{i}=\textbf{1}(x_{i1}<x_{i2})-\textbf{1}(x_{i1} \geq x_{i2}),
\end{equation*}
\begin{equation*}
P(Z_{i}=1 \mid x_{i1},x_{i2}) = \Phi(\mu_{i}),
\end{equation*}
\begin{equation*}
\alpha_{i} = 0.5*\textbf{1}(x_{i3} > 1/4) + 0.25*\textbf{1}(x_{i3} > 2/4) + 0.25*\textbf{1}(x_{i3}>3/4),
\end{equation*}
\begin{equation*}
\sigma = \frac{\theta^{(n)} - \theta^{(1)}}{8}, \ \ \ \ \theta_{i} = \mu_{i} + \alpha_{i}\Phi\left(\mu_{i}\right),
\end{equation*}
\\
\noindent
where $\theta^{(n)} = \max\left(\theta_1,...,\theta_n\right)$ and $\theta^{(1)} = \min\left(\theta_1,...,\theta_n\right)$.

\begin{figure}
	\begin{center}
		\includegraphics[scale=0.5]{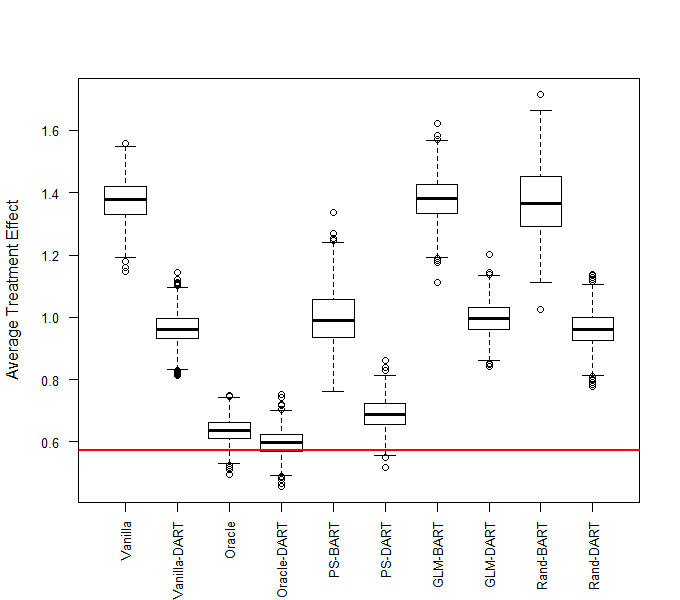}
	\end{center}
	\caption{Friedman example - Boxplots of the posterior CATE for each model from one iteration. The red line represents the true CATE. \label{fig:dart2}}
\end{figure}

The boxplots from Figure \ref{fig:dart2} are composed by posterior CATE for each model for the one replication of the simulation. The red line is the real CATE for this specific iteration. 

\begin{figure}
	\begin{center}
		\includegraphics[scale=0.5]{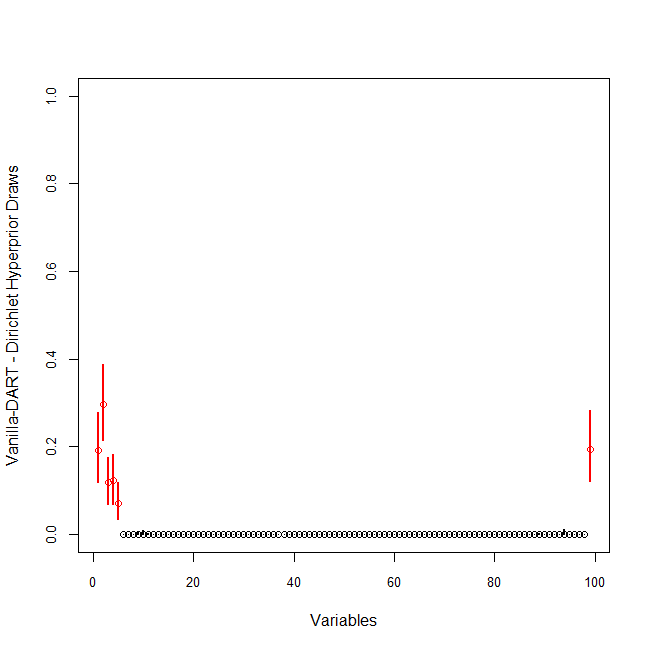}
	\end{center}
	\caption{Friedman example - Posterior draws from Vanilla-DART Dirichlet hyperprior with 95\% credible intervals.  In red: $x_1$, $x_2$, $x_3$, $x_4$, $x_5$, $\pi(\tilde{x})$ and $z$, respectively. \label{fig:vsb0}}
\end{figure}

Like in Section \ref{subsubsec:hahnsparse}, the DART models were superior. Again Vanilla and Rand models estimates are, apparently biased due to the RIC phenomenon, while the PS models have shown more consistent results in relation to the GLM models. The most accurate results were held by the Oracle models.

\begin{figure}
	\begin{center}
		\includegraphics[scale=0.5]{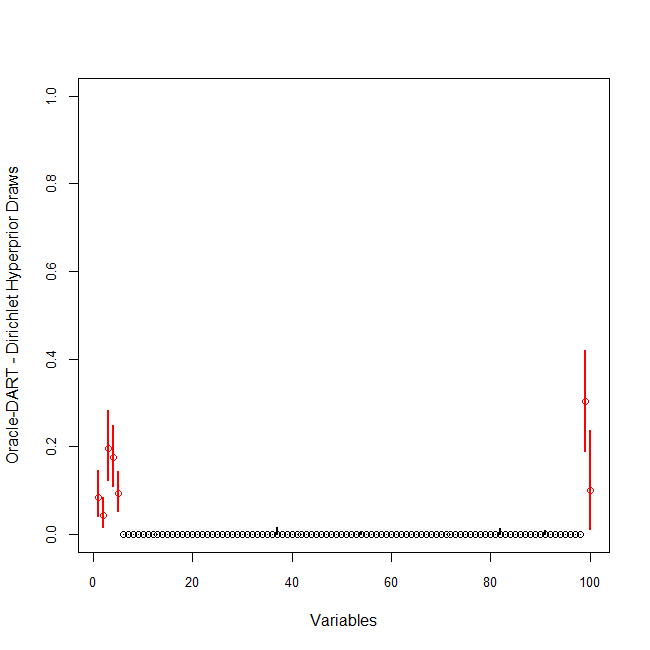}
	\end{center}
	\caption{Friedman example - Posterior draws from Oracle-DART Dirichlet hyperprior with 95\% credible intervals.  In red: $x_1$, $x_2$, $x_3$, $x_4$, $x_5$, $\pi(\tilde{x})$ and $z$, respectively. \label{fig:vsb1}}
\end{figure}

\begin{figure}
	\begin{center}
		\includegraphics[scale=0.5]{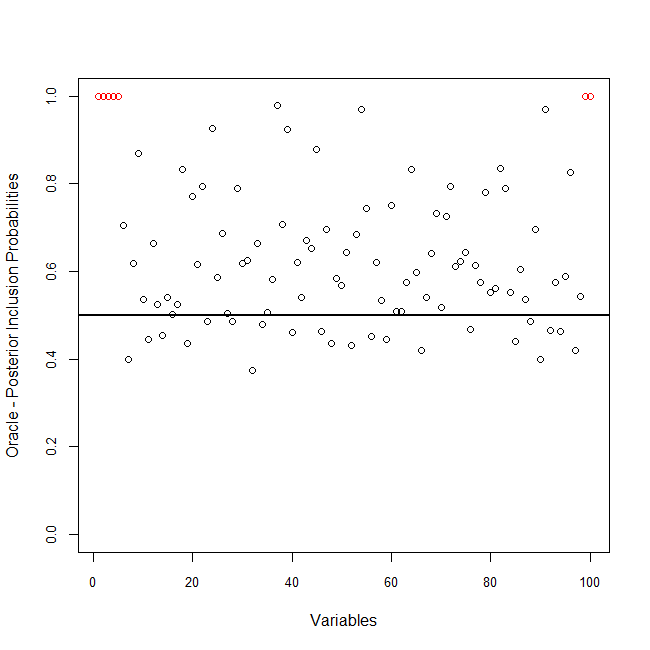}
	\end{center}
	\caption{Friedman example - Posterior Inclusion Probability of Oracle model. In red: $x_1$, $x_2$, $x_3$, $x_4$, $x_5$, $\pi(\tilde{x})$, and $z$, respectively.  \label{fig:vsb2}}
\end{figure}

\begin{figure}
	\begin{center}
		\includegraphics[scale=0.5]{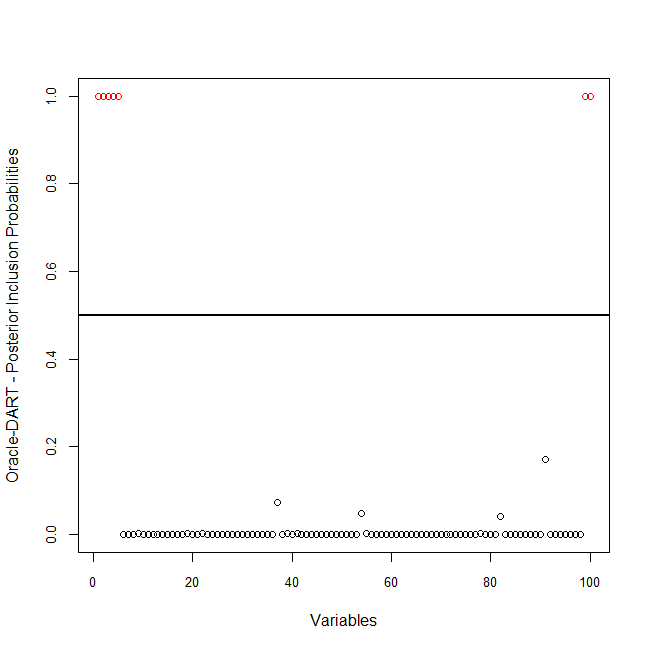}
	\end{center}
	\caption{Friedman example - Posterior Inclusion Probability of Oracle-DART model. In red: $x_1$, $x_2$, $x_3$, $x_4$, $x_5$, $\pi(\tilde{x})$, and $z$, respectively.  \label{fig:vsb3}}
\end{figure}

Figures \ref{fig:vsb0} and \ref{fig:vsb1} represents the Dirichlet hyperprior draws for the Vanilla-DART and Oracle-DART models, while the Figures \ref{fig:vsb2} and \ref{fig:vsb3} represents BART and DART $PIP$ estimation from the Oracle models. It is possible to see that the DART models tends to select only the relevant variables in the model, including the propensity score.

\begin{table}
	\centering
	\begin{tabular}{ |c|c|c|c|c|c|c|c| }
		\hline
		& Model & CATE RMSE & ITE RMSE & Precision & Recall & $F_1$ & PS-Usage \\ \hline
		\hline
		\multirow{12}{*}{BART} 
		& \multirow{2}{*}{Vanilla} & 0.675 &0.741 & 0.061 & 1.000 & 0.114 & - \\
		& & (0.109) & (0.478) & (0.000) & (0.000) & (0.001) & - \\ \cline{2-8}
		& \multirow{2}{*}{Oracle} & 0.058 & 0.354 & 0.093 & 1.000 & 0.170 & 100.000 \\
		& & (0.019) & (0.183) & (0.007) & (0.000) & (0.011) & (0.000) \\ \cline{2-8}
		& \multirow{2}{*}{PS} & 0.344 & 0.448 & 0.070 & 1.000 & 0.132 & 100.000 \\
		& & (0.100) & (0.310) & (0.001) & (0.000) & (0.002) & (0.000) \\ \cline{2-8} 
		& \multirow{2}{*}{GLM} & 0.624 & 0.692 & 0.070 & 1.000 & 0.131 & 99.845 \\
		& & (0.109) & (0.518) & (0.001) & (0.000) & (0.001) & (0.012) \\ \cline{2-8}
		& \multirow{2}{*}{Rand} & 0.668 & 0.755 & 0.060 & 1.000 & 0.113 & 77.338 \\
		& & (0.113) & (0.541) & (0.000) & (0.000) & (0.001) & (0.117) \\ \cline{2-8}
		& \multirow{2}{*}{Probit} & - & - & 0.095 & 1.000 & 0.173 & - \\
		& & - & - & (0.012) & (0.000) & (0.020) & - \\ \hline
		\hline
		\multirow{12}{*}{DART} 
		& \multirow{2}{*}{Vanilla} & 0.345 & 0.643 & 0.893 & 1.000 & 0.939 & - \\
		& & (0.107) & (0.505) & (0.116) & (0.000) & (0.068) & - \\ \cline{2-8}
		& \multirow{2}{*}{Oracle} & 0.052 & 0.375 & 0.984 & 1.000 & 0.992 & 100.000 \\
		& & (0.018) & (0.201) & (0.047) & (0.000) & (0.025) & (0.000) \\ \cline{2-8}
		& \multirow{2}{*}{PS} & 0.131 & 0.493 & 0.828 & 1.000 & 0.900 & 100.000 \\
		& & (0.060) & (0.369) & (0.137) & (0.000) & (0.085) & (0.000) \\ \cline{2-8} 
		& \multirow{2}{*}{GLM} & 0.337 & 0.589 & 0.914 & 0.910 & 0.908 & 41.195 \\
		& & (0.099) & (0.461) & (0.104) & (0.069) & (0.063) & (0.441) \\ \cline{2-8}
		& \multirow{2}{*}{Rand} & 0.344 & 0.655 & 0.765 & 0.833 & 0.794 & 0.467 \\
		& & (0.105) & (0.535) & (0.094) & (0.000) & (0.056) & (0.013) \\ \cline{2-8}
		& \multirow{2}{*}{Probit} & - & - & 0.976 & 1.000 & 0.986 & - \\
		& & - & - & (0.085) & (0.000) & (0.051) & - \\ \hline
		\hline
		
	\end{tabular}
	\caption{Friedman example - Model assessment through the means of CATE RMSE, ITE RMSE, Precision, Recall, $F_1$, and usage of the propensity score over replications. Standard deviation is given in parenthesis.}
	\label{table:friedman}
\end{table}

\begin{figure}
	\begin{center}
		\includegraphics[scale=0.5]{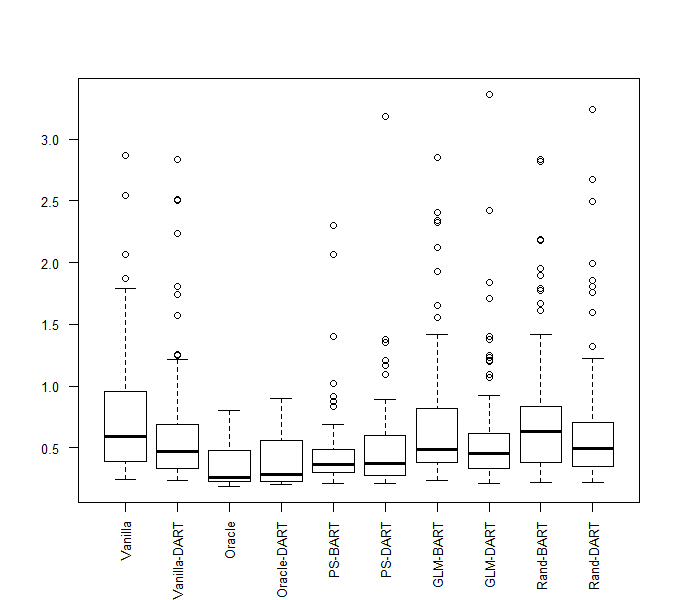}
	\end{center}
	\caption{Friedman example - Boxplots of the CATE RMSE for each model calculated over 100 simulations. \label{fig:friedcatermse}}
\end{figure}

\begin{figure}
	\begin{center}
		\includegraphics[scale=0.5]{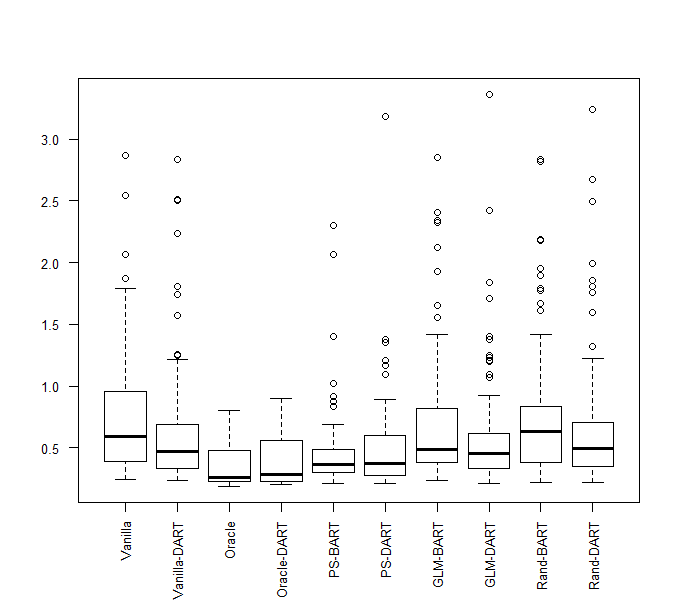}
	\end{center}
	\caption{Friedman example - Boxplots of the ITE RMSE for each model calculated over 100 simulations. \label{fig:frieditermse}}
\end{figure}

Following the Section \ref{subsubsec:hahnsparse} methodology, the performance of the models over the replications is evaluated at Table \ref{table:friedman}, along with Figures \ref{fig:friedcatermse} and \ref{fig:frieditermse}. Variables were selected via $PIP$. For Precision, Recall and $F_1$, the value $1.0$ indicates a perfect adjustment. PS-Usage indicates the mean of the proportion of times that the propensity score estimation was used in the model, and if the model had an estimation of the propensity score, regardless of misspecification, the estimation was considered a relevant variable in this analysis. All measurements are given by the mean over the replications, with standard deviation in parentheses.

\subsection{Simulations Assessment}
\label{subsec:simasses}

The simulations in Sections \ref{subsec:example} and \ref{subsec:high} were performed in order to assess the inclusion of the propensity score as a covariate and advocate for tools that can be used on treatment effect analysis.

In all examples the inclusion of the true or the estimated propensity score resulted in a decrease of the impact that the RIC phenomenon had over the model. As expected, the models which had the true propensity score have shown better results. The two models used to estimated the propensity score have shown similar performance in the simulation based on real data, but on the sparsity examples the BART based models had superior results. That may be due to the fact that the GLM was including all the variables in the model, while the BART and the DART models can naturally incorporate interactions between covariates, and even perform accurate variable selection in the case of DART. The simulations were generated in a simple setting, allowing both models to adjust relatively well, but in real datasets there might be unusual interactions between the covariates, as well as irrelevant variables, which is a scenario that models derived from BART, such as DART and BCF, can adapt with ease.

The flexibility of the ICE Plot allow it to be used under many different scenarios, but in the treatment effect setting it brings up three main advantages: allows the visualization of variables that do not impact in the treatment effect; show possible candidates of relevant variables for different individuals; and grants a way that may help in the identification of groups whose individuals may be affected in different ways by the chosen treatment.

Under the sparsity setting, it can be seen that the DART models variable selection performed well, even in both examples. Moreover, based on the results, the $PIP$ from the DART model can be considered an important tool upon the definition of which variables to include in the propensity score estimation.

\section{Discussion}
\label{sec:discuss}

In general, \cite{Hill11} and \cite{Hahn17} work were extended by including previously existing tools that could, and should, be applied to the causal inference setting.

We have corroborated \cite{Hahn17} study, which argues that inclusion of the propensity score can suppress at least part of the bias that the RIC phenomenon adds to the data. This idea was enforced by analyzing the effects of propensity score through a sensitivity analysis, and by the use of a full-Bayesian variable selection method. We have also found that in binary treatment effect observational studies even a naively estimated propensity score (which was played by the GLM in the simulations) may have a positive impact on the model, and even if the estimates are completely random (like in the Rand models in the simulations), there will be no additional bias in the treatment effect estimation in relation to the Vanilla model. Alternative Bayesian or machine learning estimations of the propensity score can also be further explored.

In regard of model effectiveness, the BCF allows its priors to be allocated freely in the functions related to the prognostic effect and the treatment effect, so the model may held better results than BART if cross-validation is applied, but we have not found a clear superior model under default priors, hyperparameters and hyperpriors.

A possible extension of this study can be done by applying the DART Dirichlet hyperprior to the BCF model and verifying the model effectiveness in high dimensional data examples with $p>n$ . Also, sensitivity analysis can be done under the sparse setting.

Another possible approach could be done by inserting heteroscedastic error terms and applying \cite{Pratola17} approach.

\bigskip
\begin{center}
{\large\bf SUPPLEMENTARY MATERIAL}
\end{center}

\begin{figure*}
	\begin{center}
		\includegraphics[scale=0.25]{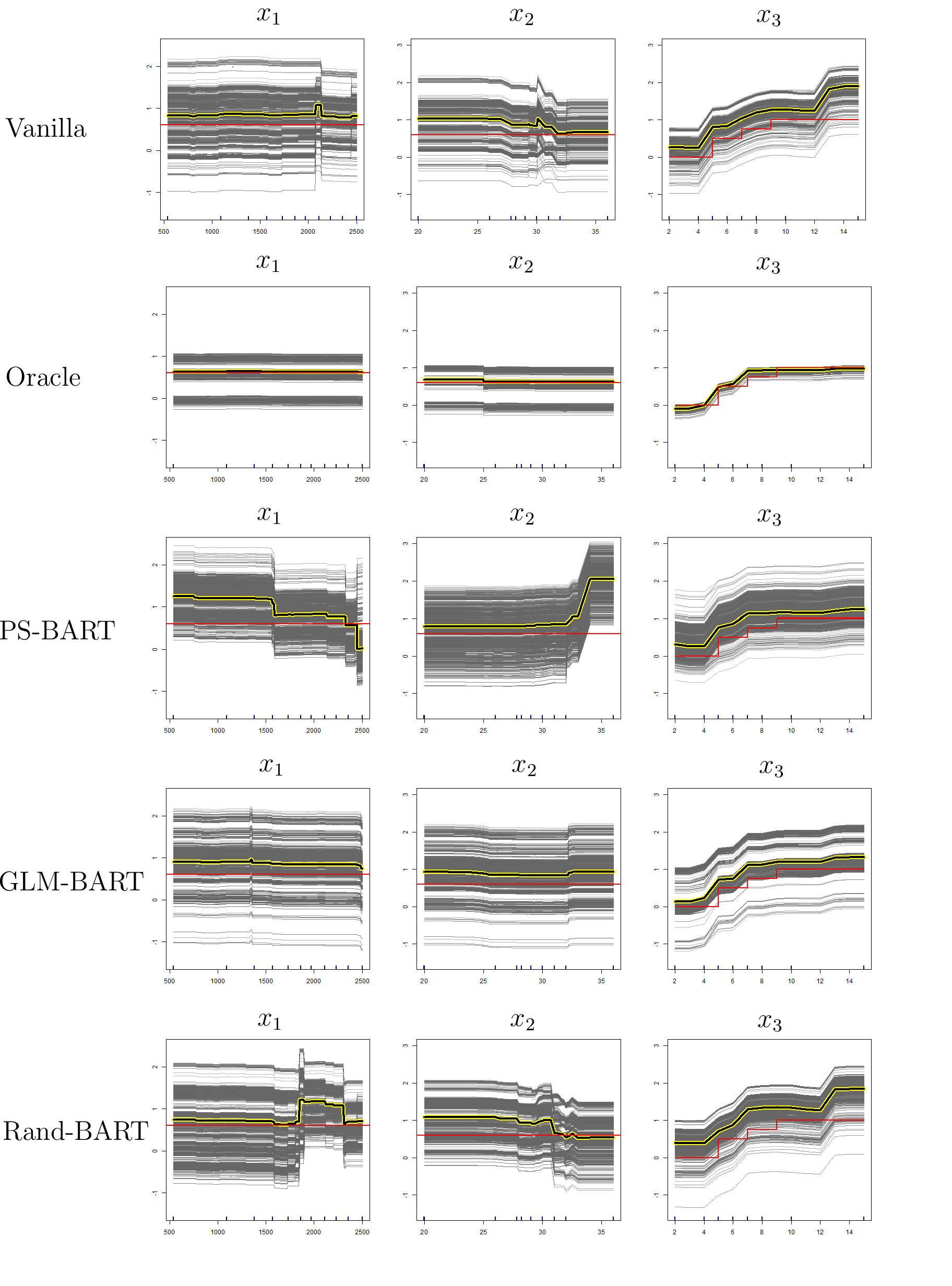}
	\end{center}
	\caption{Hill Example - ICE Plots for the impact of variables $x_1$, $x_2$, and $x_3$ on the estimated Treatment Effect. The red line represents the true average treatment effect. The black line with yellow borders is the PDP curve. The gray lines are the individual ICE curves. \label{fig:sup1}}
\end{figure*}

\begin{figure}
	\begin{center}
		\includegraphics[scale=0.25]{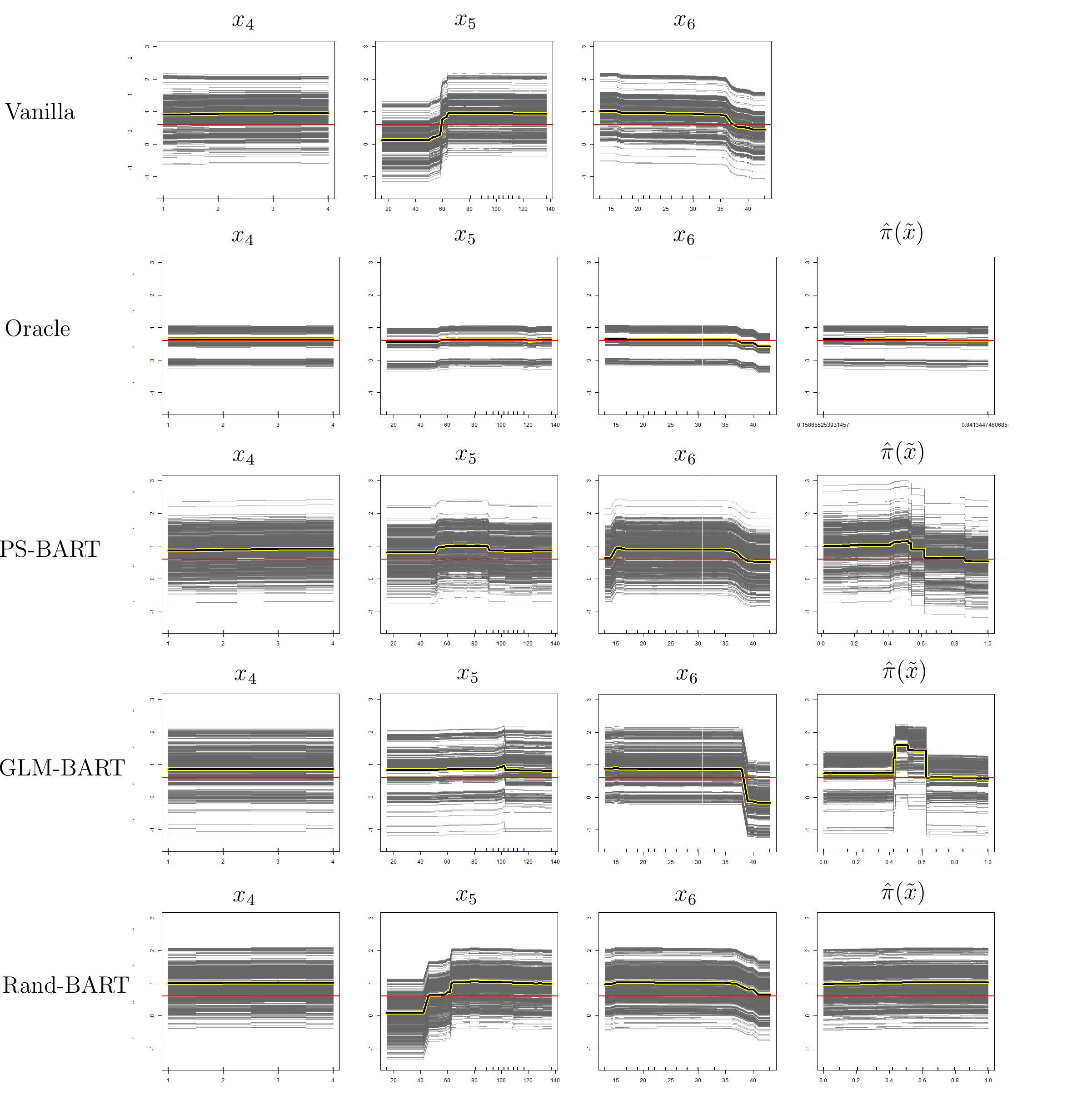}
	\end{center}
	\caption{Hill Example - ICE Plots for the impact of variables $x_4$, $x_5$, $x_6$, and $\hat{\pi}(\tilde{x})$ on the estimated Treatment Effect. The red line represents the true average treatment effect. The black line with yellow borders is the PDP curve. The gray lines are the individual ICE curves. \label{fig:sup2}}
\end{figure}

Let $y_{ij}$ denote the $i$th observation in the terminal node $i \in \left\{1,2,...,b\right\}$ in a tree with structure $T$.
It will be assumed that 
\begin{equation*}
	y_{i1},y_{i2},...,y_{in_i}\stackrel{iid}{\sim}\mathcal{N}\left(\mu_{i},\sigma^{2}\right), \ \ \ \ \mu_{1},\mu_{2},...,\mu_{b}\stackrel{iid}{\sim}\mathcal{N}\left(k,\sigma^{2}_{\mu}\right),
\end{equation*}
such that $n_1 + n_2 + ... + n_b = n$, and $M = \left\{\mu_1,...,\mu_b \right\}$.

So, the likelihood of the data following \cite{CGM98} framework is given by
\begin{equation*}
	p\left(\tilde{y} \mid \textbf{x}, T, M, \sigma^{2}\right) = \prod_{i = 1}^{b} \prod_{j = 1}^{n_{i}} \frac{1}{\sqrt{2\pi\sigma^{2}}}\exp\left(\frac{-\left(y_{ij} - \mu_{i} \right)^2}{2\sigma^{2}}\right).
\end{equation*}

In order to avoid reversible jumps, the MCMC algorithm uses 
\begin{equation*}
	p\left(\tilde{y} \mid \textbf{x}, T, \sigma \right) = \int p\left(\tilde{y} \mid \textbf{x}, T, M, \sigma^{2}\right) p\left(M \mid T, \sigma \right) dM, \ \ \ \ p\left(M \mid T, \sigma \right) = \prod_{i=1}^{b}{p\left(\mu_{i} \mid T, \sigma \right)}.
\end{equation*}

Thus, \cite{Linero17} shows that this integral can be simplified to
\begin{equation*}
p\left(\tilde{y} \mid \textbf{x}, T, \sigma \right) = \prod_{i=1}^{b}\left(2\pi\sigma^{2}\right)^{-\frac{n_{i}}{2}} \sqrt{\frac{\sigma^{2}}{n_{i}\sigma_{\mu}^{2}+\sigma^{2}}} \exp \left( -\frac{\sum_{j=1}^{n_{i}}\left(y_{ij}-\bar{y}_{i}\right)^{2} }{2\sigma^{2}} - \frac{ n_{i}\left(\bar{y}_{i} - k \right)^{2} }{ 2\left( n_{i}\sigma_{\mu}^{2} + \sigma^{2} \right) } \right),
\end{equation*}
where $\bar{y}_{i} = \frac{\sum_{j=1}^{n_{i}} y_{ij} }{n_{i}}$.

\begin{algorithm}[H]
	\SetAlgoLined
	\SetKwInOut{Input}{Input}
	\SetKwInOut{Output}{Output}
	
	\underline{function BART} $\left(\tilde{y},\textbf{x}, m, iter\right)$;\\
	\Input{Training Data with a response vector $\tilde{y}$ and covariate matrix $\textbf{x}$;\\
		 Number of trees in the ensemble - $m$;\\
		 Number of MCMC iterations - $iter$;}
	\Output{BART posterior tree draws;}
	
	Start $T_{1}, T_{2},...,T_{m}$ as single node trees;\\
	Start $M_{1}, M_{2},...,M_{m}$ (with $|M_{j}|=1$ $\forall$ $j$) filled with zeros;\\
	Start $\sigma=\hat{\sigma}_{OLS}$ as an initial guess;\\
	\For{($i$ in $1:iter$)}{
		\For{($j$ in $1:m$)}{
				Calculate $\tilde{R}_{j} \equiv \tilde{y} - \sum_{h \neq j}^{m}g(\textbf{x},T_{h},M_{h})$;\\
				Select a proposal tree $T^{*}$ from tree $T_{j}$;\\
				$T_{j}=T^{*}$ with probability $\alpha(T_{j},T^{*})=min\left\{\frac{q(T^{*},T_{j})}{q(T_{j},T^{*})}\frac{p(\tilde{R}_{j} \mid \textbf{x}, T^{*},\sigma)p(T^{*})}{p(\tilde{R}_{j} \mid \textbf{x}, T_{j},\sigma_)p(T_{j})},1\right\}$ or\\
				$T_{j}=T_{j}$ with probability $1-\alpha(T_{j},T^{*})$;\\
				Draw $M_{j}$ from $p(M_{j} \mid T_{j}, \tilde{R}_{j}, \sigma)$;\\
				
		}
		Draw $\sigma$ from $p(\sigma \mid T_{1},T_{2},...,T_{m},M_{1},M_{2},...,M_{m},\tilde{y}, \textbf{x})$;\\
		Save the $i$th posterior draw $T_{1},T_{2},...,T_{m},M_{1},M_{2},...,M_{m}, \sigma$;
	}
	
	\caption{Generating BART posterior tree draws with Bayesian backfitting}
	
\end{algorithm}

\begin{algorithm}[H]
	\SetAlgoLined
	\SetKwInOut{Input}{Input}
	\SetKwInOut{Output}{Output}
	
	\underline{function PDP \& ICE Plot} $\left(\tilde{y},\textbf{x},\hat{f}(.), i\right)$;\\
	\Input{Training Data with a response vector $\tilde{y}$ and covariate matrix $\textbf{x}={\tilde{x}_{1},...,\tilde{x}_{n}}$, with $\tilde{x}_{j}= \left\{x_{j1},x_{j2},...,x_{jp}\right\}$;\\
		   Prediction function - $\hat{f}(.)$;\\
	   	   Variable to be analyzed - $i$;}
	\Output{PDP and ICE Plot for variable $i$;}

	\For{($j$ in $1:n$)}{
			\For{($k$ in $1:n$)}{	
				Set $x_{ki}={x_{ji}}$;\\
				Calculate $\hat{f}_{ki}\left(x_{ji}\right)=\hat{f}\left(\tilde{x}_k\right)$;
			}
		Calculate $\hat{f}_{i}\left(x_{ji}\right) = \frac{1}{n}\sum_{k=1}^{n} \hat{f}_{ki}\left(x_{ji}\right)$;
	}
	
	Plotting the $n$ ICE curves:\\
	\For{($k$ in $1:n$)}{
		Plot the ICE curve relative to the $k$th individual by plotting the pairs\\
		\For{($j$ in $1:n$)}{
			Plot the pair $\left(x_{ji}\hat{f}_{ki}\left(x_{ji}\right)\right)$;
		}
	}
	
	Plot the PDP curve by plotting the pairs\\
	\For{($j$ in $1:n$)}{
		Plot the pair $\left(x_{ji}\hat{f}_{i}\left(x_{ji}\right)\right)$;
	}

	\caption{Generating PDP and ICE Plot for $|S| = 1$}
	
\end{algorithm}

\bibliographystyle{chicago}

\bibliography{Bibliography-Santos_Lopes}
\end{document}